\documentclass[sigplan,10pt,screen,balance=false]{acmart}
\pdfoutput=1
\usepackage[utf8x]{inputenc}
\usepackage[british]{babel}
\usepackage{csquotes}
\usepackage{subcaption}
\usepackage{microtype}
\usepackage{balance}

\newtoggle{anon}
\togglefalse{anon}
\newcommand{\redacted}[1]{\iftoggle{anon}{<redacted>}{#1}}

\definecolor{keywordcolor}{rgb}{0.7, 0.1, 0.1}   
\definecolor{commentcolor}{rgb}{0.4, 0.4, 0.4}   
\definecolor{symbolcolor}{rgb}{0.0, 0.1, 0.6}    
\definecolor{sortcolor}{rgb}{0.1, 0.5, 0.1}      
\definecolor{errorcolor}{rgb}{1, 0, 0}           
\definecolor{stringcolor}{rgb}{0.5, 0.3, 0.2}    

\usepackage{listings}

\lstdefinestyle{leannocolor}{
  basicstyle={\ttfamily},
  identifierstyle={\ttfamily},
  keywordstyle=[1]{\ttfamily},
  keywordstyle=[2]{\ttfamily},
  keywordstyle=[3]{\ttfamily},
  stringstyle={\ttfamily},
  commentstyle={\ttfamily}
}
\lstMakeShortInline[style=leannocolor]|

\setcopyright{acmlicensed}
\acmPrice{15.00}
\acmDOI{10.1145/3437992.3439928}
\acmYear{2021}
\copyrightyear{2021}
\acmSubmissionID{poplws21cppmain-p31-p}
\acmISBN{978-1-4503-8299-1/21/01}
\acmConference[CPP '21]{Proceedings of the 10th ACM SIGPLAN International Conference on Certified Programs and Proofs}{January 18--19, 2021}{Virtual, Denmark}
\acmBooktitle{Proceedings of the 10th ACM SIGPLAN International Conference on Certified Programs and Proofs (CPP '21), January 18--19, 2021, Virtual, Denmark}


\begin{document}

\begin{abstract}
  In theorem provers based on dependent type theory such as Coq and Lean,
  induction is a fundamental proof method and induction tactics are omnipresent
  in proof scripts. Yet the ergonomics of existing induction tactics are not
  ideal: they do not reliably support inductive predicates and relations; they
  sometimes generate overly specific or unnecessarily complex induction
  hypotheses; and they occasionally choose confusing names for the hypotheses
  they introduce.

  This paper describes a new induction tactic, implemented in Lean~3, which
  addresses these issues. The tactic is particularly suitable for educational
  use, but experts should also find it more convenient than existing induction
  tactics. In addition, the tactic serves as a moderately complex case study for
  the metaprogramming framework of Lean~3. The paper describes some difficulties
  encountered during the implementation and suggests improvements to the
  framework.
\end{abstract}

\author{Jannis Limperg}
\affiliation{%
  \institution{Vrije Universiteit Amsterdam}
  \department{Department of Computer Science}
  \streetaddress{NU Building, De Boelelaan 1111}
  \city{Amsterdam}
  \postcode{1081 HV}
  \country{The Netherlands}}
\email{j.b.limperg@vu.nl}
\orcid{0000-0002-8861-5231}
\title{A Novice-Friendly Induction Tactic for Lean}
\date{15 December 2020}

\maketitle

\section{Introduction}\label{sec:intro}

Induction is a fundamental proof method in dependently typed interactive theorem
proving, and so in proof assistants such as Coq and Lean, the induction tactic
is usually among the first tactics a novice encounters. Yet despite this
elevated position, current induction tactics exhibit a number of usability
issue. Experts have got used to these dark corners and routinely perform the
busywork required to get around them. Newcomers to dependently typed theorem
proving, however, often struggle to connect the workings of the induction tactic
with their informal understanding of proof by induction. \redacted{Jasmin
  Blanchette} and colleagues noticed this when designing the \redacted{Logical
  Verification} course at \redacted{Vrije Universiteit Amsterdam}, which uses
Lean to teach interactive theorem proving. They had to devote significant space
and time to technical issues with Lean's standard induction tactic, which
distracted from more fundamental topics.

To shield novices (and experts) from these distractions, we need induction tactics
which minimise the gap between formal and informal proof by induction. This
paper describes an attempt at such a tactic. In particular, it addresses three
usability issues with existing induction tactics.

\paragraph{Indexed inductive types.} The standard induction tactics of Coq and
Lean sometimes produce counterintuitive goals when used with indexed inductive
types. Consider the type |fin|~|n| of natural numbers strictly less than |n|,
which can be encoded as an indexed inductive type. Given this encoding, it is
trivial on paper to prove by induction (or even mere case distinction) that
|fin|~|0| is uninhabited. But when we apply the standard induction tactics of
Coq and Lean to this goal, they produce a goal which asks us, unhelpfully, to
prove |false| in an empty context. Similar problems occur regularly when
formalising programming language metatheory, which tends to use indexed
inductive types extensively to define inductive predicates and relations.

Coq and Lean both provide case splitting tactics which prove that |fin|~|0| is
uninhabited, but which cannot deal with goals that require induction. While
this is obviously useful, it is pedagogically unfortunate: case splitting should
produce the same goals as induction, only without the induction hypotheses. This
connection is obscured if case splitting and induction tactics handle indexed
types differently.

Coq (but not Lean) also provides |dependent| |induction|, an alternative induction
tactic which handles indexed inductive types better. It proves our lemma about
|fin|~|0| immediately. But this tactic creates a new issue: it often produces
unnecessarily complex induction hypotheses. In one of the examples I discuss
below, |dependent| |induction| creates the induction hypothesis
\begin{lstlisting}
∀ S' t',
  (while (λ _, true) S, t) =
  (while (λ _, true) S', t') →
  false
\end{lstlisting}
which is obviously equivalent to |false|.

My induction tactic works like |dependent| |induction|, but simplifies the
induction hypotheses to eliminate redundant arguments. This combination yields
intuitive goals in many situations involving indexed induction types.

\paragraph{Overly specific induction hypotheses.} Standard induction tactics
tend to produce overly specific induction hypotheses. Consider this humble
injectivity lemma:
\begin{lstlisting}
∀ (n m : ℕ), n + n = m + m → n = m
\end{lstlisting}
If we perform induction on |n| after introducing |n|, |m| and the equation, we
get an induction hypothesis about a fixed |m|:
\begin{lstlisting}
n + n = m + m → n = m
\end{lstlisting}
Unfortunately, this induction hypothesis does not help us make progress with the
proof. Instead, we need an induction hypothesis which generalises over all
|m|:
\begin{lstlisting}
∀ m, n + n = m + m → n = m
\end{lstlisting}

Convincing Coq or Lean to generate the more general induction hypothesis is not
hard. We can introduce |m| after the induction (rather than before) or use
special syntax offered by the standard induction tactics. However, this still
presents a problem for novices: they must first recognise that the original
induction hypothesis is not helpful and that generalising |m| is the appropriate
remedy. Neither of these insights will be obvious to someone not yet familiar
with the mechanics of interactive theorem proving.

That is why my induction tactic does the opposite of the standard tactics:
rather than generating the most specific induction hypotheses by default, it
generates the most general ones. In effect, it generalises every hypothesis like
we generalised |m|, with some restrictions to avoid obviously useless
generalisations. This sometimes produces overly general induction hypotheses,
but for novices, that is a better problem to have. They can simply specialise an
induction hypothesis to recover the more specific version.

\paragraph{Naming.} Induction tactics generate many new hypotheses and these
hypotheses must be named. This may seem like a trivial concern, but Lean's
standard induction tactic shows that it is not. Consider this simple lemma about
a particular formulation of the transitive closure |tc|~|r| of a binary relation
|r| (which in Lean would usually live in the universe |Prop| of propositions,
but for simplicity I pretend throughout that there is only one universe |Type|):
\begin{lstlisting}
∀ α (r : α → α → Type) (a b c : α)
  (h₁ : tc r a b) (h₂ : tc r b c), tc r a c
\end{lstlisting}
When we use Lean's standard induction tactic to perform induction on
|h₁|, it generates a rather forbidding goal:
\begin{lstlisting}
α : Type
r : α → α → Type
c a b h₁_x h₁_y h₁_z : α
h₁_hr : r h₁_x h₁_y
h₁_ht : tc r h₁_y h₁_z
h₁_ih : tc r h₁_z c → tc r h₁_y c
h₂ : tc r h₁_z c
⊢ tc r h₁_x c
\end{lstlisting}

The new hypotheses' names are clearly too cumbersome: experts and novices alike
will want to immediately rename almost everything. Worse, the names are
misleading because the first element of the transitive chain, |a|, is now |h₁_x|
while |b| has become |h₁_z|. Novices will struggle to make the connection
between the old and new hypotheses, and thus to understand how this goal is
connected to the lemma they wanted to prove. Coq's standard induction tactic
generates more helpful names than Lean's, but it too misses this connection.

To prevent the resulting confusion, my induction tactic uses a number of
heuristics to generate names that reflect common intuitions about how induction
works. For the |tc| example, it generates this goal:
\begin{lstlisting}
α : Type
r : α → α → Type
a y b c : α
hr : r a y
h₁ : tc r y b
ih : ∀ c, tc r b c → tc r y c
h₂ : tc r b c
⊢ tc r a c
\end{lstlisting}
Omitting the |h₁| prefixes and preserving the names |a|, |b| and |c| makes for a
much more reasonable-looking goal.

\medskip

The new induction tactic, which addresses the above usability issues, is
described in Sect.~\ref{sec:impl}, after I lay some terminological groundwork in
Sect.~\ref{sec:induction}. The description proceeds chronologically through
every step the new induction tactic takes when processing a goal.

The new induction tactic could be implemented in any dependently typed proof
assistant with indexed inductive types and a suitable metaprogramming framework.
(Parts of it do, however, make use of the somewhat controversial axiom~K.) An
implementation for Lean~3 is available in mathlib~\cite{mathlib}, Lean's de
facto standard library.\footnote{The version of mathlib which contains the
  implementation described in this paper is available online at
  \url{https://github.com/leanprover-community/mathlib/tree/d36af184d154f2e99f60fec5cd71bb3e53899d5c}.
  The relevant source file is \texttt{src/\allowbreak tactic/\allowbreak
    induction.lean}. An archive of this code is available at
  \url{https://doi.org/10.5281/zenodo.4327209}.} The tactic was also used in the
2020 edition of the \redacted{Logical Verification} course, where it replaced
Lean's standard induction tactic. This allowed the course authors to
significantly simplify the lecture notes and accompanying code, since the tricks
experts use to make Lean's standard induction tactic work did not need to be
taught any more.

The new induction tactic is among the larger tactics written in Lean's
metaprogramming framework~\cite{ebner2017}. This provides an opportunity to
evaluate how the framework fares on a moderately complex task. In
Sect.~\ref{sec:meta}, I describe some problems I encountered while implementing
the tactic, as well as possible workarounds and suggestions for improvements.
This case study will hopefully be useful to aspiring Lean metaprogrammers and
may guide the design of the metaprogramming framework in the upcoming fourth
version of Lean.

In summary, I make the following contributions:
\begin{itemize}
  \item I describe an induction tactic that is more ergonomic than the state of
    the art. The design is geared particularly towards novice users,
    but experts should also find it easier to work with.
  \item I provide an implementation of this tactic in Lean~3, which previously
    lacked a convenient induction tactic.
  \item I give an experience report about Lean's metaprogramming framework,
    pointing out some pitfalls and suggesting improvements.
\end{itemize}

\section{Induction in Dependent Type Theory}\label{sec:induction}

Induction in dependent type theories is intimately connected with indexed
inductive types~\cite{dybjer1994}, a fundamental concept of most modern
dependently typed theorem provers. Indexed inductive types generalise
non-indexed inductive types such as natural numbers and lists and are often used
to define inductive predicates and relations.

A typical example of this use is the transitive closure of a binary relation,
which can be encoded in Lean as follows:
\begin{lstlisting}
inductive tc {α : Type} (r : α → α → Type) :
  α → α → Type
| base : ∀ x y (hr : r x y), tc x y
| step : ∀ x y z (hr : r x y) (ht : tc y z),
    tc x z
\end{lstlisting}
The above defines a type family |tc| of type
\begin{lstlisting}
∀ {α : Type} (r : α → α → Type),
  α → α → Type
\end{lstlisting}
The first argument of |tc|, |α|, will be left implicit, as
indicated by the curly braces. The second argument is the relation |r|
whose transitive closure we are taking. The third and fourth arguments are
elements of |α| that are related by |tc|~|r|.

The transitive closure is inductively generated by two rules corresponding to
the two constructors of |tc|. The |base| constructor says that if two elements
|x| and |y| are related by |r|, then they are also related by |tc|~|r|. The
|step| constructor says that if |r| relates |x| and |y|, and |tc|~|r| relates
|y| and |z|, then |tc|~|r| relates |x| and |z|.

In the type of |tc|, we distinguish between two kinds of arguments:
\emph{parameters} and \emph{indices}. Arguments that appear before the colon,
here |α| and |r|, are parameters of |tc|.
Parameters are implicitly quantified over in the types of |tc|'s
constructors, and whenever |tc| appears in a constructor type, it is
implicitly applied to the parameters. Thus, the full type of
|base| is
\begin{lstlisting}
∀ {α} {r : α → α → Type} (x y : α),
  r x y → tc r x y
\end{lstlisting}

The arguments of |tc| after the colon are its indices. Unlike parameters, these
may vary freely in the constructor types, and indeed our constructors
instantiate the indices of |tc| with different expressions.

Each inductive type has an associated induction principle, the (dependent)
\emph{recursor}, which reflects the fact that every closed element of the
inductive type consists of finitely many constructor applications. In Lean, a
recursor is added as an axiom whenever we define an inductive
type~\cite{carneiromsc}. For |tc|, we get the recursor |tc.rec|, whose type
appears in Fig.~\ref{fig:tc.rec}. This type is derived from |tc| as follows:
\begin{itemize}
  \item The first two arguments, |α| and |r|, are the
    parameters of |tc|.
  \item |M| is the type we are constructing, also known as the
    \emph{motive} of the induction. It is a predicate over elements of
    |tc| (and its indices).
  \item |Base| and |Step| are \emph{minor premises} corresponding to the
    constructors of |tc|. They ask users to give one proof of |M| for the case
    where |tc|~|r|~|x|~|y| was proved by |base| and one for the case where
    |tc|~|r|~|x|~|y| was proved by |step|. In the |step| case, we may assume a
    proof of |M| for the recursive constructor argument |ht|.
  \item From all this data, |tc.rec| concludes |M|~|x|~|y|~|e| for an arbitrary
    element |e| of |tc|~|r|~|x|~|y|. We call |e| the \emph{major premise} of the
    induction. This is the hypothesis on which we perform induction.
\end{itemize}

\begin{figure}
\begin{tabular}{c}
\begin{lstlisting}
∀ α (r : α → α → Type)
  (M : ∀ x y, tc r x y → Type)
  (Base: ∀ x y (hr : r x y),
    M x y (base x y hr))
  (Step : ∀ x y z (hr : r x y)
    (ht : tc r y z),
    M y z ht →
    M x z (step x y z hr ht))
  x y (e : tc r x y),
  M x y e
\end{lstlisting}
\end{tabular}
\caption{The type of \texttt{tc.rec}}\label{fig:tc.rec}
\end{figure}

To perform induction in Lean, then, means to apply the recursor of an inductive
type. For an example, we return to the transitivity of the transitive closure:

\noindent
\begin{minipage}{\linewidth}
\begin{lstlisting}
∀ α (r : α → α → Type) (x y z : α)
  (hxy : tc r x y) (hyz : tc r y z), tc r x z
\end{lstlisting}
\end{minipage}

We first fix |α|, |r|, |x|, |y| and |z|. The proof then proceeds by recursion on
|hxy|, so this is our major premise. The parameters, |α| and |r|, are already
determined by this choice, as are the major premise indices |x| and |y|. For the
motive, we choose
\begin{lstlisting}
M := λ (x y : α) (_ : tc r x y),
       tc r y z → tc r x z
\end{lstlisting}
Substituting this motive in the minor premises, we are left with one proof
obligation for each constructor, corresponding to the cases of the induction.
The proof of our lemma then reads:
\begin{lstlisting}
λ α r x y z (hxy : tc r x y),
  tc.rec α r
    (λ x y _, tc r y z → tc r x z)
    <proof of Base minor premise>
    <proof of Step minor premise>
    x y hxy
\end{lstlisting}

An induction tactic helps with this rather arduous exercise by automating much
of it. Ideally, users do not have to contend with motives, parameters or indices
and are presented only with one intuitive new goal for each constructor.
The next section explains how to achieve this in many cases.

\section{Implementation of the Induction Tactic}\label{sec:impl}

The following subsections describe each step the new induction tactic takes to
perform an induction, in chronological order.

\subsection{Generalisation of Complex Indices}\label{sec:index-generalisation}

The first problem our induction tactic must solve is the treatment of complex
index arguments in a major premise. To see what this means, consider a typical
task in programming language metatheory: a simple lemma about the big-step
semantics of a toy imperative language.

The abstract syntax of our toy language is defined by the (non-indexed)
inductive type |stmt| in Fig.~\ref{fig:stmt}. Its constructors
represent, from top to bottom, a no-op statement; variable assignment;
sequencing of statements; and a while loop. The |state| type mentioned
by some constructors represents the current program heap as a map from variable
names to their current values (which, for simplicity, are always natural
numbers). The loop condition of a while loop is given as a predicate on the heap
state.

\begin{figure}
\begin{tabular}{c}
\begin{lstlisting}
inductive stmt : Type
| skip   : stmt
| assign : string → (state → ℕ) → stmt
| seq    : stmt → stmt → stmt
| while  : (state → Type) → stmt → stmt
\end{lstlisting}
\end{tabular}
\caption{Syntax of a toy imperative language}\label{fig:stmt}
\end{figure}

The language's big-step semantics are given by the indexed inductive type
|big_step| in Fig.~\ref{fig:big_step}, omitting some constructors. This type
defines a relation between a program |S|, an initial state |s| and a final state
|t|. If |big_step| |(S,|~|s)|~|t| is derivable, then |S|, when executed in heap
state |s|, terminates in heap state |t|. We write |(S,|~|s)|~|⇒|~|t| for
|big_step| |(S,|~|s)|~|t|. To enable this notation in Lean~3, and following
standard informal practice, the first argument of |big_step| is a pair type (so
|big_step| is partially uncurried).

\begin{figure}
\begin{tabular}{c}
\begin{lstlisting}
inductive big_step :
  stmt × state → state → Type
| skip {s} :
  big_step (skip, s) s
| while_true {b : state → Type} {S s t u}
    (hcond : b s)
    (hbody : big_step (S, s) t)
    (hrest : big_step (while b S, t) u) :
  big_step (while b S, s) L
| while_false {b : state → Type} {S s}
    (hcond : ¬ b s) :
  big_step (while b S, s) s
| ...
\end{lstlisting}
\end{tabular}
\caption{Big-step semantics of the toy language}\label{fig:big_step}
\end{figure}

Now we want to prove that the infinite loop does not terminate. In Lean, this
means solving the following goal:
\begin{lstlisting}
S : stmt
s t : state
h : (while (λ _, true) S, s) ⇒ t
⊢ false
\end{lstlisting}
Above the turnstile appear the local hypotheses of our goal, most importantly
|h|, which says that the infinite loop steps to some state
|t| and thus terminates. Right of the turnstile is our target, the
canonical empty type |false|. On paper, this goal is easily proven by
induction on the derivation of |h|. Lean and Coq's default induction
tactics, however, fail us. Applying them yields unprovable subgoals.

This is because in the type of |h|, |big_step| has a
\emph{complex} index. A term is complex if it is anything other than a local
hypothesis. If such a term appears as an index of an inductive type, trouble
ensues. Here, the offending complex index is the first argument of
|big_step|,
\begin{lstlisting}
(while (λ _, true) S, s)
\end{lstlisting}

A naive induction tactic now proceeds as follows. Our target is
|false|, which depends neither on the hypothesis |h| nor its
indices, so the motive of the induction is the constant function
\begin{lstlisting}
M := λ (x : stmt × state) (t : state)
       (p : x ⇒ t), false
\end{lstlisting}
Constructing the type of |big_step|'s recursor according to the schema
from Sect.~\ref{sec:induction}, we get the following minor premise for the
|skip| constructor:
\begin{lstlisting}
∀ (s : state), M (skip, s) s skip
\end{lstlisting}
Yet this gives a plainly unprovable goal if we substitute the motive
|M|:
\begin{lstlisting}
∀ (s : state), false
\end{lstlisting}
The root cause of this issue is that by applying the recursor like we did, we
effectively forgot that the first index of the major premise involved a
|while|, not a |skip|. As a result, we cannot recognise that
the major premise could not have been constructed by |big_step|'s
|skip| constructor.

This deficiency of induction tactics in the presence of complex indices is well
known. The traditional solution, in the context of dependent type theory, is due
to McBride~\cite{mcbride2002}. The remainder of this section describes a variant
of his procedure, with one major change. McBride's tactic analyses arbitrary
elimination principles, determines which of their arguments lead to problems
similar to our complex index problems and generalises those arguments. In
contrast, we only support the standard recursors of inductive types, for which
we know that issues like the one we have seen are only caused by complex
indices. This makes our tactic less general, but it also considerably simplifies
the implementation (and its presentation below). In an educational setting,
where custom elimination principles are rarely used, this seems like an
acceptable trade-off. Coq's |dependent| |induction| uses a very similar
restricted variant of McBride's approach (which is, to my knowledge, not
described in the literature). Coq does support custom elimination principles
with |dependent| |induction|, but in this case the tactic still only generalises
complex indices. This capability could also be added, with moderate engineering
effort, to our tactic.

McBride's solution to the complex index problem is to replace any complex index
|i| with a new hypothesis |Hi|, called an \emph{index placeholder}, and to add
an \emph{index equation} |Hi|~|=|~|i| to the target. This ensures that we do not
lose information about the value of the index. Applying this transformation
yields an equivalent goal:
\begin{lstlisting}
S : stmt
s t : state
Hi : stmt × state
h : Hi ⇒ t
⊢ Hi = (while (λ _, true) S, s) → false
\end{lstlisting}

Then we proceed as before. But since our goal now depends on the first index of
|h|, we generate a different motive for the induction:
\begin{lstlisting}
λ (x : stmt × state) (t : state) _,
  x = (while (λ _, true) S, s) → false
\end{lstlisting}
The minor premise for |skip| changes accordingly, leaving us with this goal for
the |skip| case:
\begin{lstlisting}
S : stmt
s s' : state
ieq : (skip, s') = (while (λ _, true) S, s)
⊢ false
\end{lstlisting}
This goal is provable because the equation |ieq|, which is derived from the index
equation, is contradictory. We have, in effect, remembered that the index of
|big_step| was a |while|, not a |skip|.

To make this index generalisation procedure work for more complex goals, we must
address two technical complications. First, when we replace a complex index in
the major premise, we generally want to also replace it in the target and in the
types of other hypotheses, to make sure that the goal remains type-correct. This
is a somewhat crude heuristic since the replacement may itself introduce type
errors, but it is right more often than wrong. However, we never replace the
index in hypotheses that occur in the type of the major premise.

A second complication arises when there are dependencies between the indices of
an inductive family. Consider, for example, the family
\begin{lstlisting}
F : ∀ (x : X) (y : Y x), Type
\end{lstlisting}
where |x| and |y| are indices, and suppose that we want to perform induction on
the hypothesis |h|~|:|~|F|~|t|~|u| (with |t| and |u| complex terms). The index
generalisation procedure then replaces |t| with a new hypothesis |Ht|~|:|~|X|
such that |Ht|~|=|~|t| and |u| with a new hypothesis |Hu|~|:|~|Y|~|Ht| such that
|Hu|~|=|~|u|. But this last equation is not well-typed: |u| has type |Y|~|t|, not
|Y|~|Ht|. In this situation, we must use a \emph{heterogeneous} equation, written
|Hu|~|==|~|u|, where the two sides may have different types.

At this point, one might become concerned for novice users of the induction
tactic: would they not get overwhelmed with index placeholders and index
equations? Fortunately, Sect.~\ref{sec:index-unification} shows that the new
hypotheses can usually be eliminated automatically after we have applied the
recursor, so our users do not get to see them.

\subsection{Generalisation of Induction Hypotheses}\label{sec:ih-generalisation}

One of the more arcane aspects of Coq and Lean's existing induction tactics is
that they ask their users to specify which hypotheses can vary during the
induction and which are fixed. This leads to counterintuitive behaviour even in
simple cases. Pierce~\cite{sf1} illustrates the problem with the following
injectivity lemma:
\begin{lstlisting}
∀ (n m : ℕ), n + n = m + m → n = m
\end{lstlisting}
An unsuspecting novice will mechanically introduce |n|, |m| and the equation,
then perform induction on |n|. This produces the following goal for the
successor case:
\begin{lstlisting}
n m : ℕ
ih : n + n = m + m → n = m
h : n + n + 2 = m + m
⊢ n + 1 = m
\end{lstlisting}
Unfortunately, this gives us an induction hypothesis, |ih|, that is not
applicable: it presumes |n|~|=|~|m| when we have, according to |h|, |n| =
|m|~|-|~|1|. The solution is to let |m| vary during the induction instead of
keeping it fixed, which gives a more sensible goal:
\begin{lstlisting}
n m : ℕ
ih : ∀ m, n + n = m + m → n = m
h : n + n + 2 = m + m
⊢ n + 1 = m
\end{lstlisting}
Now |ih| can be instantiated with |m|~|-|~|1| to close the goal.

Generalising the induction hypothesis in this manner is not difficult, and
existing tactics provide special syntax for it. But to a novice, it will be far
from obvious that this is why our first proof attempt gets stuck. Novices often
have trouble recognising that a goal is unprovable in the first place, and when
they do, they may suspect any number of errors on their part. A novice-friendly
induction tactic should therefore not fix every hypothesis by default, as the
existing tactics do, but rather \emph{generalise} every hypothesis. This
sometimes leads to an overly general induction hypothesis, but that is much less
harmful: the user does not get stuck but merely has to apply the induction
hypothesis to some additional arguments. Our new tactic also offers a convenient
syntax to fix some or all hypotheses; if all hypotheses are fixed, the tactic
behaves like the existing induction tactics.

Implementing this automatic generalisation is straightforward in most cases. We
simply revert (\enquote*{unintroduce}) all hypotheses before applying the
recursor. However, there are three classes of hypotheses that should not be
reverted:
\begin{enumerate}
  \item Hypotheses which the user has explicitly fixed and their dependencies,
    i.e.\ those hypotheses which occur in the type of a fixed hypothesis. If we
    were to revert a dependency of a fixed hypothesis, we would also have to
    revert the fixed hypothesis.
  \item Hypotheses on which the major premise depends. Such hypotheses cannot be
    reverted without also reverting the major premise.
  \item Hypotheses which would not make the induction hypotheses more general if
    we were to revert them.
\end{enumerate}

The last class deserves further analysis. Usually, when we perform an induction,
all hypotheses are relevant to the proof, so generalising a hypothesis leads to
a more general induction hypothesis. However, that is not always the case. In
longer proofs, it is occasionally convenient to prove a helper lemma (by
induction) inline, without leaving the proof environment. These lemmas may
involve only some of the hypotheses, but if we follow our generalise-everything
approach, we also generalise all the other hypotheses in the context which have
nothing to do with the helper lemma. This gives us an induction hypothesis with
additional redundant arguments.

Consider this example:
\begin{lstlisting}
x : X
n m : ℕ
⊢ n + m = m + n
\end{lstlisting}
The first hypothesis, |x|, has nothing to do with the rest of the goal ---
perhaps we were in the middle of a proof involving |x| and decided to prove
commutativity of addition inline as a helper lemma. To that end, we perform
induction on |n|. The naive generalisation algorithm would now revert both |x|
and |m|, yielding an induction hypothesis with an obviously redundant argument:
\begin{lstlisting}
∀ (x : X) (m : ℕ), n + m = m + n
\end{lstlisting}

To prevent this, we revert a hypothesis |h| only if it meets at least one of the
following criteria:
\begin{enumerate}
  \item |h| occurs in the target. Recall that the motive of the induction, which
    determines the induction hypothesis, is derived from the target. So if the
    target has the form |∀|~|x,|~|P|~|x| instead of |P|~|h|, we get a more general
    induction hypothesis. In the commutativity example, |m| fulfils this
    criterion, so it is generalised.
  \item |h| depends on the major premise, or on any of the dependencies of the
    major premise. Then, |h| is likely to be a property of the major premise
    that is relevant to the induction. For instance, if we have a major premise
    |n|~|:|~|ℕ|, a hypothesis |h|~|:| |n|~|>|~|0| and a target |P n|, the motive
    of the induction should be derived from the generalised target
    |n|~|>|~|0|~|→| |P|~|n|. Otherwise we would not be able to use the fact that
    |n|~|>|~|0| during the induction, and in particular we would not be able to
    discharge the case for |n|~|=|~|0| by noting that |0|~|≯|~|0|. The same reasoning
    also applies to dependencies of the major premise: performing the induction
    may give us additional information about these dependencies, so hypotheses
    mentioning them may be of interest.
\end{enumerate}

Conversely, any hypothesis |h| that does not meet either criterion --- such as
|x| in the commutativity example --- should not be generalised. Such hypotheses
have no connection to either the target or the major premise, so generalising
them only adds redundant arguments to the induction hypothesis. Of course, our
criteria only prevent the most obvious forms of over-generalisation: for the
commutativity proof, |m| does not need to be generalised either.

\medskip

This step concludes the preprocessing, so now our tactic applies the recursor.
To do so, we would usually have to generate a motive, which involves solving a
higher-order unification problem. Luckily, Lean has a built-in heuristic that
generates correct motives most of the time, so we do not have to concern us with
this issue here.

By applying the recursor, we generate one new goal for each case of the
induction (i.e.\ each minor premise). The next steps are applied to each of these
goals individually.

\subsection{Unification of Index Equations}\label{sec:index-unification}

In Sect.~\ref{sec:index-generalisation}, we introduced placeholders for the
complex indices of the major premise and index equations to remember what the
placeholders stand for. We can usually eliminate these equations again after the
recursor has been applied, using McBride's |Qnify| tactic~\cite{mcbride1996}.

|Qnify| implements a form of first-order unification. It works on a queue
of equations which initially contains the equations for each index, starting
with the first. The order is important when indices depend on each other since
unification of earlier index equations may simplify later ones. The two sides of
each equation are unified by applying the following set of rules until no rule
applies any more:

\paragraph{Substitution.} For an equation |eq|~|:| |x|~|=|~|t| or |eq|~|:|
|t|~|=|~|x|, where |x| is a local hypothesis and |t| is a term in which |x| does
not occur, delete |eq| and replace |x| with |t| everywhere in the goal.

\paragraph{Injection.} For |eq|~|:| |C|~|t₁|~|...| |tₙ|~|=| |C|~|u₁|~|...| |uₙ|,
where |C| is a constructor of an inductive type, delete |eq| and add new
equations |tᵢ|~|=|~|uᵢ|. The new equations are added to the front of the queue,
so they are processed immediately after this step. Some of the equations may
have to be heterogeneous.

\paragraph{Conflict.} For |eq|~|:| |C|~|t₁|~|...| |tₙ|~|=| |D|~|u₁|~|...| |uₘ|,
where |C| and |D| are distinct constructors, solve the goal since |eq| is
contradictory.

\paragraph{Deletion.} For |eq|~|:| |t|~|=|~|u|, where |t| and |u| are
definitionally equal, delete |eq|.

\paragraph{Cycle.} For |eq|~|:| |x|~|=|~|t| (or symmetric), where |x| appears
under constructors in |t|, solve the goal since |eq| is contradictory. The
previous condition means that |t| must be of the form
\begin{lstlisting}
C₁ ... (C₂ ... (Cₙ ... x ...) ...) ...
\end{lstlisting}
where the |Cᵢ| are all constructors of the same inductive type and |n| is not
zero. For example, this rule would match the equation |x|~|=|
|succ|~|(succ|~|(succ|~|x))|, where |succ| is the successor constructor of |ℕ|.

\paragraph{Homogenisation.} For |eq|~|:| |t|~|==|~|u|, where |t|~|:|~|T|,
|u|~|:|~|U| and |T| is definitionally equal to |U|, replace |eq| with the
equivalent homogeneous equation |t|~|=|~|u|. This rule typically applies because
the types |T| and |U| were initially distinct --- hence the heterogeneous
equation |t|~|==|~|u| --- but became definitionally equal during the unification
of earlier equations.

\medskip

Recall the example from Sect.~\ref{sec:index-generalisation}. After generalising
complex indices, we ended up with this goal in one of the cases of the induction:
\begin{lstlisting}
S : stmt
s s' : state
ieq : (skip, s') = (while (λ _, true) S, s)
⊢ false
\end{lstlisting}
Applying |Qnify| to the index equation |ieq|, we first use the injection rule
since both sides of the equation are applications of the pair constructor
|(_,_)|. This gives us new equations:
\begin{lstlisting}
ieq₁ : skip = while (λ _, true) S
ieq₂ : s' = s
\end{lstlisting}
We then apply the conflict rule to |ieq₁| since |skip| and |while| are different
constructors, solving the goal. Thus, users of our tactic never get to see this
case of the induction.

The homogenisation rule, which deals with heterogeneous equations, is only valid
in certain type theories, namely those in which Streicher's
axiom~K~\cite{hofmann1994} is derivable. This includes Lean, but excludes some
other popular proof assistants, particularly those that seek to be compatible
with the univalence axiom of homotopy type theory~\cite{hottbook}, which is
incompatible with axiom~K. Induction tactics for such type theories would not
use McBride's index generalisation method but rather that of Cockx et
al.~\cite{cockx2014, cockx2018}, who show how to achieve a similar effect
without using axiom K.

Implementing the unification procedure is straightforward except for the cycle
rule. To prove that an equation
\begin{lstlisting}
eq : x = C₁ (... (Cₙ x) ...)
\end{lstlisting}
is contradictory, we use a size measure |sizeof| which counts the number of
constructors in a term. Lean generates this measure for every inductive type.
Applying it on both sides of the equation yields an equation in |ℕ|:
\begin{lstlisting}
eq : sizeof x = sizeof x + n
\end{lstlisting}
For positive |n|, this can be discharged by applying an appropriate lemma.

\subsection{Simplification of Induction Hypotheses}\label{sec:ih-simp}

The index placeholders and index equations introduced in
Sect.~\ref{sec:index-generalisation} also occur as additional arguments to the
induction hypotheses generated by the recursor application. But like the
equations themselves, these arguments can often be trivially eliminated. Prior
work does not address this issue: Coq's |dependent| |induction| tactic makes no
attempt to simplify the induction hypotheses and Lean's |cases| tactic, which
also uses McBride's index generalisation technique, does not generate induction
hypotheses in the first place.

Let us again consider the |big_step| example from
Sect.~\ref{sec:index-generalisation}, but now we focus on the first inductive
case. After the index equations have been eliminated, we get the goal shown in
Fig.~\ref{fig:before-ih-simp}, corresponding to the |while_true| constructor of
|big_step|.
\begin{figure*}
\begin{tabular}{c}
\begin{lstlisting}
S : stmt
s t u : state
h₁ : (S, s) ⇒ t
h₂ : (while (λ _, true) S, t) ⇒ u
ih₁ : ∀ S' t', (S, s) = (while (λ _, true) S', t') → false
ih₂ : ∀ S' t', (while (λ _, true) S, t) = (while (λ _, true) S', t') → false
⊢ false
\end{lstlisting}
\end{tabular}
\caption{A goal before simplification of induction hypotheses}\label{fig:before-ih-simp}
\end{figure*}

The induction hypotheses, |ih₁| and |ih₂|, have been generalised
over two index placeholders, |S'| and |t'|, and an index equation.
But in |ih₂|, these are all redundant. We can only hope to apply
|ih₂| if we instantiate |S'| with |S| and |t'| with
|t|; any other instantiation (modulo propositional equality) would not
satisfy the index equation.

This is a common case, so we postprocess the induction hypotheses to eliminate
such redundant arguments. To do so, we first replace each index placeholder in
the type of |ih₂| with a fresh metavariable:
\begin{lstlisting}
(while (λ _, true) S, t) =
(while (λ _, true) ?S', ?t') →
false
\end{lstlisting}

We then iterate through the index equations --- here only one --- and unify the
left-hand side of each with the right-hand side, using Lean's built-in
unification procedure. If unification finds a unique solution, the
metavariables are assigned accordingly:
\begin{lstlisting}
?S' := S
?t' := t
\end{lstlisting}

Now we specialise the induction hypothesis, applying it to the terms we assigned
to the metavariables:
\begin{lstlisting}
(while (λ _, true) S, t) =
(while (λ _, true) S, t) →
false
\end{lstlisting}

Finally, we delete any index equation whose left-hand side is definitionally
equal to its right-hand side. This leaves us with a pleasantly simple induction
hypothesis:
\begin{lstlisting}
ih₂ : false
\end{lstlisting}

This procedure does not always succeed in eliminating the index placeholders and
index equations. If we apply it to the first induction hypothesis,
|ih₁|, it instantiates the |t'| index placeholder with
|s|, but it does not find a unique solution for |S'|. The
induction hypothesis thus remains unwieldy:
\begin{lstlisting}
ih₁ : ∀ S',
  (S, s) = (while (λ _, true) S', s) → false
\end{lstlisting}
This is pedagogically unfortunate, as students are unlikely to fully understand
why an equation appears in |ih₁|. Simplifying the equation to eliminate the |s|
on both sides would help a little, but I have not encountered enough such
situations in practice to justify complicating the implementation.

Besides redundant index equations, an induction hypothesis can also contain
contradictory index equations, e.g. \verb|skip|~\verb|=| |while|~|b|~|S|. The induction
hypothesis can then never be applied and should be deleted. Unfortunately, my
induction tactic currently does not do this, due to a limitation of Lean's
built-in unification procedure, which does not allow us to distinguish between
terms that are certainly unequal, such as |skip| and |while|~|b|~|S|, and terms
that might be propositionally equal, such as |S| and |while|~|b|~|?S'|. The
|Qnify| procedure from Sect.~\ref{sec:index-unification} could be adapted to
this use case.

\subsection{Naming of Constructor Arguments}\label{sec:naming}

A surprisingly large portion of the new induction tactic is dedicated to naming.
Finding intuitive names is important, particularly in an educational setting.
When the names are chosen poorly, novices (and occasionally experts) may have
trouble understanding how the new goals relate to the old goal. Lean's standard
induction tactic uses a simple, predictable naming scheme, but the generated
names are plainly too cumbersome for use in education. Consider again the fact
that the transitive closure operator from Sect.~\ref{sec:induction} is
transitive:
\begin{lstlisting}
∀ α (r : α → a → Type) (a b c :α)
  (h₁ : tc r a b) (h₂ : tc r b c), tc r a c
\end{lstlisting}

Performing induction on |h₁|, Lean's induction tactic generates a rather
intimidating goal in the inductive case, shown in
Fig.~\ref{fig:naming:lean-old}. The goal illustrates a number of common
problems:

\begin{figure*}
  \centering
  \begin{subfigure}{.3\textwidth}
    \begin{lstlisting}
    α : Type
    r : α → α → Type
    c a b h₁_x h₁_y h₁_z : α
    h₁_hr : r h₁_x h₁_y
    h₁_ht : tc r h₁_y h₁_z
    h₁_ih : tc r h₁_z c →
      tc r h₁_y c
    h₂ : tc r h₁_z c
    ⊢ tc r h₁_x c
    \end{lstlisting}
    \caption{Lean's standard induction tactic}\label{fig:naming:lean-old}
  \end{subfigure}
  \begin{subfigure}{.3\textwidth}
    \begin{lstlisting}
    α : Type
    r : α -> α -> Type
    c x y z : α
    hr : r x y
    h₁ : tc r y z
    IHh₁ : tc r z c ->
      tc r y c
    h₂ : tc r z c
    ⊢ tc r x c
    \end{lstlisting}
    \caption{Coq's standard induction tactic}\label{fig:naming:coq}
  \end{subfigure}
  \begin{subfigure}{.3\textwidth}
    \begin{lstlisting}
    α : Type
    r : α → α → Type
    a y b c : α
    hr : r a y
    h₁ : tc r y b
    ih : ∀ c, tc r b c →
      tc r y c
    h₂ : tc r b c
    ⊢ tc r a c
    \end{lstlisting}
    \caption{The new induction tactic}\label{fig:naming:lean-new}
  \end{subfigure}
  \caption{Goals for the same proof produced by three different induction tactics}\label{fig:naming}
\end{figure*}

\begin{itemize}
  \item All new hypotheses generated by the induction tactic are prefixed with
    |h₁|. This clarifies their origin, but it also makes the goal hard
    to understand at a glance. Names like |h₁_x| are simply too long
    compared to a plain |x|.
  \item The first and middle elements of the transitive chain, which were called
    |a| and |b| in the lemma statement, are now called
    |x| and |z| (disregarding the |h₁| prefix).
    One could hardly blame a novice for being confused about how the old and new
    hypotheses relate to each other.
  \item As if to make the previous problem worse, Lean's standard induction
    tactic does not remove the hypotheses |a| and |b|, even
    though they are now redundant and have been effectively replaced by
    |h₁_x| and |h₁_z|.
\end{itemize}

Coq's standard induction tactic fares better, producing the goal in
Fig.~\ref{fig:naming:coq}. The tactic drops the |h₁| prefixes and correctly
removes the redundant hypotheses. Yet it, too, renames |a| to |x| and |b| to
|z|.

The new induction tactic fixes this last issue, producing the goal in
Fig.~\ref{fig:naming:lean-new}. It recognises the connection between old and new
hypotheses and names the new ones accordingly. The name |y| is not ideal, but
other than that, no name would be out of place in an informal proof.

To achieve this effect, we employ the following algorithm. Suppose we are in the
case of the induction corresponding to a constructor |C|. Then we need to
name the following new hypotheses: one hypothesis for each of |C|'s
arguments; one induction hypothesis for each of |C|'s recursive
arguments; and any index placeholders and index equations we have introduced and
not subsequently eliminated.

For the last category, a simple schema suffices. Index placeholders and index
equations are usually eliminated anyway, but if they remain in the goal, we
name them |index_i| and |induction_eq_i| for some |i|.

Naming the induction hypotheses is also fairly straightforward. If there is only
a single induction hypothesis, we name it |ih|. Otherwise, we use
names like |ih_e|, where |e| is the hypothesis to which this
induction hypothesis applies (meaning the hypothesis corresponding to the
recursive constructor argument which gives rise to |ih_e|). For
example, if we perform induction on some expression type, we may get
subexpressions |e₁| and |e₂| and induction hypotheses
|ih_e₁| and |ih_e₂|. Coq uses a similar scheme. Lean's
standard induction tactic simply numbers the generated induction hypotheses,
which is usually less helpful.

Finally, we consider the constructor arguments, where the naming problem becomes
interesting. Suppose we want to name the hypothesis corresponding to an argument
|a|~|:|~|A| of constructor |C|. Then we try each rule from the following list,
stopping at the first one that applies. These heuristics seem to yield intuitive
names in many cases --- though humans use so many different heuristics that
trying to incorporate them all would be a fool's errand. When our heuristics
fail, users can of course give their own names.

In larger developments, one should usually give explicit names anyway to make
the proof script more robust. Still, having the induction tactic generate
sensible names makes for a better user experience in the experimental phase of
proof development, when the proof script is not yet polished.

\paragraph{Recursion.} If |a| is a recursive argument, it is named after the
major premise. So if we eliminate a natural number |n|, the number in the
inductive case is also called |n|; if we eliminate an expression |e|, its
subexpressions are called |e|, |e_1|, etc. These are likely to be good names
since the subexpressions are of the same type as the parent expression. In
Fig.~\ref{fig:naming:lean-new}, |h₁| is derived from a recursive argument in
this way. Coq also uses this rule.

\paragraph{Index association.} If |a| is associated with an index
argument, it is named after that index argument. This is the rule responsible
for our improvement over Coq in the example from Fig.~\ref{fig:naming}. We say
that the argument |x| of |tc|'s |step| constructor
is associated with the first index of |tc|. In the hypothesis
|h₁|, which we are performing induction on, that first index is
instantiated with |a|, so the hypothesis corresponding to
|x| is named |a|.

Capturing this situation in general requires a somewhat involved criterion.
Suppose we are naming an argument \verb|a|~\verb|:|~\verb|A| of a constructor
|C| whose return type is |F|~|j₁|~|...|~|jₙ|, where |F| is an inductive family
with |n| indices. We say that |a| is associated with the |i|th index if it
occurs in |jᵢ|. Now suppose our major premise is |e|~|:| |F|~|k₁|~|...|~|kₙ|.
Consider those |kᵢ| such that |a| is associated with the |i|th index. If these
|kᵢ| are all the same hypothesis |h|, and if the type of |h| is definitionally
equal to the type of |a|, then the hypothesis corresponding to |a| is named
after |h|.

The stipulation about definitionally equal types exists to prevent confusion
when a constructor argument is associated with an index of a different type. In
such cases, it is usually better not to name the argument after the index, since
names are often related to the types of the named entities. For instance, if an
argument |a|~|:|~|α| is associated with an index |as|~|:|~|list α|, we do not
want the hypothesis corresponding to |a| to be called |as|. The restriction
could perhaps be relaxed to allow, for instance, an argument of type |list|~|α|
to be associated with an index of type |list|~|β|, but I have found no need for
this in practice.

\paragraph{Named arguments.} If |a| is named in the definition of
the constructor |C|, that name is used. In our example, the
|Step| constructor has an argument called |hr|, so the
corresponding hypothesis is also called |hr|. Coq also uses this rule.

\paragraph{Type-based naming.} If |a|'s type, |A|, is associated with a list of
typical variable names, we use these. Such an association is given by an
instance of the type class |variable_names| for |A|, which contains a list of
names. Later, when the tactic looks for a name for |a|, it performs a type class
instance search for |variable_names|~|A|. If it finds an instance, it uses the
first unused name from the associated list. We give such instances for some
standard types (associating, for example, the names |n| and |m| with the type
|ℕ|), but users can override these with their own higher-priority instances. The
type class mechanism also allows us to give natural names for data structures
such as lists: if a type |A| is associated with the variable name |x|,
|list|~|A| is by default associated with the name |xs|. This mechanism was
developed for the new induction tactic, but can be used by other tactics as
well.

\paragraph{Fallback.} If none of the above rules apply, |a| receives a
default name: |h| if |A| is a proposition and |x| otherwise.

\medskip

The first three rules are ordered somewhat arbitrarily. I have found the given
order to be the one that most often matches common naming preferences, but there
are many examples where a different order would fit better. The example from
Fig.~\ref{fig:naming} would arguably be improved if |h₁| was called
|ht| instead, using the name from the constructor declaration rather
than the recursion rule. But switching the priority of these rules would also
change other names for the worse.

If the name chosen for |a|, say |n|, is already in use, we fall back to |n_1|,
|n_2|, etc. This complicates the implementation because many of the induction
tactic's processing steps may remove hypotheses, so we only know which names are
in use after all steps have finished. To address this issue, we initially give
the introduced hypotheses temporary names, then run the naming algorithm as the
last step of the tactic to obtain the final names.

\section{Evaluation of Lean's Metaprogramming Framework}\label{sec:meta}

I have implemented the tactic described in the previous section in the
metaprogramming framework~\cite{ebner2017} of Lean~3. This provides an
opportunity to evaluate how the framework fares on a moderately complex task.

\subsection{Overview of the Framework}\label{sec:meta-overview}

Like other modern metaprogramming approaches such as Mtac2~\cite{kaiser2018} or
Idris's elaborator reflection~\cite{christiansen2016}, Lean metaprograms are
written in Lean itself rather than its implementation language C++. They are
marked with the |meta| keyword, which signifies a stage separation:
|meta| definitions may refer to non-|meta| ones, but not the other
way around. Metaprograms can therefore be inconsistent (e.g.\ they need not
terminate) without compromising the consistency of the non-|meta|
fragment. At the same time, metaprograms have access to all the data structures
and functions defined in non-|meta| Lean, avoiding duplicate effort.

Most metaprograms are \emph{tactics}, which means they have type |tactic|~|α|
for some |α|. The |tactic| type family is a Haskell-style monad which provides
an imperative embedded domain-specific language for writing tactics. Tactics
operate on a tactic state with zero or more goals. A goal has a local context,
containing the current list of hypotheses, and a target type; the objective of a
tactic is usually to construct an element of the target type (represented as an
abstract syntax tree). To do this, tactics can make use of a large number of
built-in tactics which manipulate hypotheses and the target, add and remove
goals, query and add definitions, unify expressions, check whether two
expressions are definitionally equal, and more.

This framework generally works well and leads to a remarkably seamless
integration between regular programs and metaprograms. Still, while implementing
the new induction tactic, I encountered some situations where it was less
helpful or clear than it could be. The next subsections discuss these cases,
which will hopefully be useful to prospective Lean metaprogrammers as well as
designers of similar metaprogramming systems.

\subsection{Tracking of Hypotheses}\label{sec:tracking}

As mentioned, most tactics operate within a local context containing the
hypotheses that are currently available. Internally, these hypotheses are
represented as expressions identified by a unique name. They also have an
external name which is shown to the user and is not necessarily unique in the
context.

Many tactics manipulate the context in some way, e.g.\ by adding or removing
hypotheses or changing the types of existing hypotheses. The trouble with this
is that any such modification changes the unique names of any affected
hypotheses. As a result, any expression involving the changed hypotheses becomes
invalid: it refers to a hypothesis that, to Lean, does not exist any more.

As an example, consider the unification procedure from
Sect.~\ref{sec:index-unification}. It operates on a queue of index equations,
unifying each in turn. Naturally, we would want to represent this queue as a
list of expressions, with each expression identifying one equation hypothesis.
But this does not work. Unifying the first equation may change the types of
subsequent equations and thus their unique names. When we then turn to the next
equation in the queue, Lean will rightfully point out that the context contains
no hypothesis with that unique name. The entire tail of the queue has been
potentially invalidated.

I encountered this issue multiple times --- it occurs whenever one needs to track
hypotheses across calls to potentially context-altering tactics, which are
numerous and do not always document the fact that they may invalidate
hypotheses.

One could imagine various workarounds for this issue. For the unification
procedure, I ended up identifying hypotheses not by unique name but by external
name. For this to work, the external names must be unique in the context, which
in this case can be ensured since the induction tactic controls these names.
This workaround is less applicable when dealing with preexisting hypotheses,
whose external names may not be unique. Another possible approach would be to
have context-altering tactics report a mapping from old unique names to new
unique names, which would allow callers to update any stored expressions. This
would, however, require changes to many tactics and callers would still have to
manually perform the update.

Perhaps the most convenient solution to this issue would be the introduction of
yet another name for hypotheses: a \emph{stable name} which would remain
unchanged when a hypothesis is modified. This would most closely reflect the
tactic writer's intuition that changing the type of a hypothesis does not make
it a different hypothesis.

\subsection{Definitional Equality}\label{sec:defeq}

Any author of tactics for a dependently typed proof assistant must contend with
definitional equality: different expressions that are equal up to computation.
For instance, |ℕ| and |let|~|T|~|:=| |ℕ|~|in|~|T| are definitionally
equal types. Many tactics should treat them as interchangeable, though this
depends on the tactics' use cases and user expectations. Checking for
definitional equality, which involves partially normalising expressions, carries
a sometimes considerable performance cost, so it would be too much to ask for a
metaprogramming framework that fully abstracts over definitional equality.

Still, Lean additionally complicates the matter in two ways. First, it lacks a
comprehensive programming interface for pattern-matching on expressions up to
definitional equality. Tactic authors must manually normalise expressions as
much as necessary, using relatively rudimentary normalisation tactics, if they
want to take definitional equality into account. Novice metaprogrammers can
hardly be expected to do this accurately, and experts may be tempted to cut
corners and ignore the issue. This shifts the burden onto tactic users, who need
to make sure that their goals have just the right shape. While implementing the
new induction tactic, I added the beginnings of an up-to-definitional-equality
matching framework to mathlib, but so far this covers only a few constructions.

The second way in which Lean complicates definitional equality is by introducing
an additional notion of transparency. Each definition is marked with one of
several transparency values, which indicate how eagerly the definition should be
unfolded during normalisation. This is reasonable, and perhaps necessary: some
definitions should indeed be unfolded almost always, others almost never.

However, the programming interface around transparency encourages mistakes. Most
tactics which take a transparency argument make this argument optional, so if
one does not supply an explicit transparency, a default value is used. This
makes it easy to make subtle mistakes (and I have made a few) where transparency
is not propagated or the wrong transparency is used. It does not help that
different tactics have different default transparency values.

\subsection{Elaboration}\label{sec:elaboration}

When writing a tactic, one must often construct expressions of some specific
form. Lean provides essentially two ways to do this: directly, by writing out
the abstract syntax tree of an expression (perhaps as a syntactically more
pleasant quotation), or by elaborating a pre-expression.

Pre-expressions are an abstract syntax representation of the expressions that
users write in Lean's surface syntax. They are turned into regular expressions
in a process called elaboration, which fills in many details --- mainly implicit
and instance arguments --- that users may thankfully omit.

Lean also allows us to use elaboration in tactics, which can be convenient. When
writing expressions directly, we have to fill in many implicit arguments as well
as universe parameters, a somewhat obscure feature of the type theory that one
would usually prefer not to think about. Elaboration can do this for us, making
tactics more readable and maintainable since they need only specify the main
parts of an expression.

Unfortunately, Lean's programming interface again makes using elaboration more
difficult than it needs to be. This is mostly due to easily avoidable limitations:
many functions that construct or deconstruct expressions operate only on fully
elaborated expressions, even though they could also work with pre-expressions.
As a result, Lean encourages its users to elaborate early, before an expression
is fully constructed. But then the elaboration algorithm lacks information about
the context in which a partially constructed expression will be used, so it can
infer less implicit arguments. Due to these limitations, I have usually found it
more convenient to write out the fully elaborated expression after all.

\subsection{Nested and Mutual Inductive Types}\label{sec:ginductive}

Lean takes the dictum that a proof system's kernel should be as small as
possible more seriously than most dependently typed theorem provers. One
ramification of this philosophy is that Lean's kernel does not support nested
and mutual (collectively: generalised) inductive types. Instead, when a user
writes a generalised inductive type, Lean compiles it to an equivalent
non-generalised inductive type during elaboration. This approach makes the
kernel smaller and thus more trustworthy. However, Lean's implementation also
illustrates a major disadvantage: hiding the internal compilation process from
users of the system requires much engineering effort.

Lean does this imperfectly and as a result, generalised inductive types are a
leaky abstraction. In metaprograms, the abstraction is in fact nonexistent:
metaprograms only get to see the internal representation of a generalised
inductive type. Thus, if a metaprogram wants to, for example, report an error
about a particular generalised inductive type to the user, it has to
reverse-engineer that generalised inductive type from its internal
representation.

This illustrates a more general issue with tactics which act on the kernel
language rather than the source language: details about the user-level program
invariably get lost in translation. The induction tactic suffers from this in a
small way. One of the naming rules from Sect.~\ref{sec:naming} checks whether a
constructor argument is named in the definition of the constructor. But in the
kernel language, all arguments are named, so if a user writes the constructor
type |X|~|→|~|Y|, our tactic sees |∀|~|(a|~|:|~|X),|~|Y|. Thus, when the tactic
encounters a nondependent argument with name |a| (or |a_1| etc.), it assumes
that this argument was not explicitly named --- but that assumption can be
mistaken. Perhaps the user really wrote |∀|~|(a|~|:|~|X),|~|Y| in the hope that
our naming rule would pick up the argument name |a|. (The community edition of
Lean~3 recently changed the elaborator so that the default name for an unnamed
argument is not |a| but $\breve{\alpha}$, assuming that no user would choose that
character as a variable name. This mitigates the issue, but $\breve{\alpha}$ now
occasionally shows up in goals to confuse Lean neophytes.)

Perhaps the most effective way to prevent such loss of information would be to
associate to each kernel expression the surface expression from which it was
elaborated. For an inductive type, this would be the |inductive|
declaration the user wrote; for a top-level definition, the equations given to
the equation compiler. In the extreme, elaboration would become reversible,
so tactics would be able to reconstruct the full program text.

Lean's treatment of generalised inductive types in metaprograms also illustrates
another issue. Effectively the only primitive metaprogram that is aware of
generalised inductive types is a normalisation procedure, users of which can
choose whether constructors of generalised inductive types should be unfolded to
their internal representation. As it happens, this is sufficient for the
purposes of our induction tactic. But it shows how a feature that was supposed
to be dealt with during elaboration still permeates large parts of the system:
every tactic that uses normalisation must decide what to do about constructors
of generalised inductive types. Given such complications, one might wonder
whether it would have been preferable to put generalised inductive types in the
kernel language after all.

\subsection{Open Expressions}\label{sec:open}

While the previous sections have been critical of some parts of the
metaprogramming framework, this section discusses a reasonable design choice
that may nevertheless be surprising to novice tactic writers: the handling of
open expressions. An expression is open when it contains at least one free
variable. Such expressions occur naturally when we deconstruct terms with
binders. For example, consider the following type:
\begin{lstlisting}
∀ (n : ℕ) (f : fin ℕ), P n f
\end{lstlisting}
We can deconstruct this type into argument types |ℕ| and |fin|~\verb|#0| and
result type |P|~\verb|#1|~\verb|#0|. The \verb|#0| and \verb|#1| are free
variables, represented as De Bruijn indices, which refer to the variable bound
by, respectively, the first and second preceding binder.

Lean lets us construct such open expressions, but it does not let us to do much
with them since most built-in tactics only work on closed expressions. Open
expressions cannot, for example, be type-checked or unified.

Instead, Lean's metaprogramming framework encourages users to treat expressions
as \emph{locally nameless}~\cite{mcbride2004}. This means we effectively use
hypotheses as free variables: while deconstructing an expression, we immediately
replace any free variables with fresh hypotheses of the appropriate type. Our
above example, so deconstructed, has argument types |ℕ| and |fin|~|cn| and
result type |P|~|cn|~|cf|, where |cn|~|:|~|ℕ| and |cf|~|:| |fin|~|cn| are fresh
hypotheses.

This representation is considerably easier to work with, not only because Lean
prefers it but also because we do not have to track the contexts of each
expression as closely. Observe, for example, that in the first decomposition of
our example, the \verb|#0| in the second argument and the \verb|#0| in the
result type refer to different arguments. The locally nameless representation
avoids such confusion.

There is one downside to this representation: Lean makes no particular effort to
optimise the construction and deconstruction of locally nameless expressions, so
these operations can be somewhat inefficient.

\section{Conclusion}\label{sec:conclusion}

I have shown how to build a user-friendly induction tactic which is particularly
suited to an educational setting. The tactic liberates its users from some of
the technical, nonessential difficulties with existing induction tactics. It
automatically generalises complex indices, ensuring that information contained
in the indices of a hypothesis is not lost. It simplifies the resulting
induction hypotheses, which would otherwise be obscured by redundant arguments.
It automatically generalises induction hypotheses as much as possible so that
users do not get stuck with an overly specific induction hypothesis. And it uses
various heuristics to generate suitable names for all the new hypotheses
it introduces. These usability improvements may spare experts some of the
tedium of pre- and postprocessing their goals, and they should lift a
considerable cognitive burden from novices.

I have also discussed some issues with Lean's metaprogramming framework which I
encountered while implementing the new induction tactic. Some of these are
easily fixable (but impactful) limitations of the programming interface; others
point to deeper issues with aspects of the framework's design. I hope that this
discussion will help make Lean's already pleasant metaprogramming even better.

\begin{acks}
  Jasmin Blanchette helped determine what features a novice-friendly induction
  tactic should have, provided many test cases and commented in great detail on
  drafts of this paper. Anne Baanen, Floris van Doorn, Gabriel Ebner, Rob Lewis
  and the anonymous reviewers gave detailed and insightful feedback on drafts of
  this paper and on the underlying code. The Lean Zulip community, particularly
  Mario Carneiro and Gabriel Ebner, patiently answered my many questions about
  Lean metaprogramming. Many thanks!

  This project was funded by the \grantsponsor{NWO}{NWO}{https://www.nwo.nl/en}
  under the Vidi programme (project No.~\grantnum{NWO}{016.Vidi.189.037}, Lean
  Forward).
\end{acks}

\balance

\bibliographystyle{ACM-Reference-Format}
\bibliography{lit}

\end{document}